\documentclass{elsart}  
\journal{Carbon}

\usepackage[pdftex]{graphicx}
\usepackage{pslatex}
\usepackage{amssymb,amsmath}  
\usepackage{textcomp}  

\usepackage[numbers]{natbib}
\usepackage[pdftex, linkbordercolor={1 1 1}]{hyperref}

\hypersetup{
pdfauthor = {E. Einarsson, Y. Murakami, M. Kadowaki, S. Maruyama},
pdftitle = {Growth dynamics of vertically aligned single-walled carbon nanotubes from in situ measurements},
}

\begin{document}
\begin{frontmatter}

\title{Growth dynamics of vertically aligned single-walled carbon nanotubes from {\em in situ\/} measurements}

\author[Todai]{Erik Einarsson},
\author[Todai,Rice]{Yoichi Murakami}, 
\author[Todai]{Masayuki Kadowaki},
\author[Todai]{Shigeo Maruyama\corauthref{cor}}
\corauth[cor]{Corresponding author Tel/Fax: +81-3-5800-6983}
\ead{maruyama@photon.t.u-tokyo.ac.jp}

\address[Todai]{Department of Mechanical Engineering, The University of Tokyo, 
\\Tokyo 113-8656, Japan}
\address[Rice]{Department of Electrical and Computer Engineering, Rice University, 
\\Houston, TX 77005-1892, USA}

\begin{abstract}
An {\em in situ\/} optical absorbance measurement was used to study the growth dynamics of vertically aligned single-walled carbon nanotubes (VA-SWCNTs) synthesized by chemical vapor deposition of ethanol. The growth rate of the VA-SWCNT film was found to decay exponentially from an initial maximum, resulting in an effective growth time of approximately 15 minutes. Investigation of various growth conditions revealed an optimum pressure at which growth is maximized, and this pressure depends on the growth temperature. Below this optimum pressure the synthesis reaction is first-order, and the rate-limiting step is the arrival of ethanol at the catalyst. We also present a novel method for determining the burning temperature of low-mass materials, which combines the {\em in situ\/} absorbance measurement with controlled oxidation.
\end{abstract}


\end{frontmatter}

\section{Introduction}
The potential to utilize the various exceptional properties~\cite{Saito-text,Dresselhaus-text} of single-walled carbon nanotubes (SWCNTs) in a wide range of novel applications~\cite{Saito-text,Dresselhaus-text,Baughman-Science297} has motivated much research in nanoscience and nanotechnology. However, the realization of most proposed applications requires not only scalable methods of synthesizing high-purity SWCNTs, but also control over their location and orientation. A significant development in this area was the synthesis of vertically aligned (VA-)SWCNT films, which was first achieved~\cite{Murakami-CPL385} by combining a dip-coat catalyst loading process~\cite{Murakami-CPL377} with the alcohol catalytic chemical vapor deposition (ACCVD) method~\cite{Maruyama-CPL360,Murakami-CPL374}. Since then, several groups have reported a variety of different methods for VA-SWCNT synthesis, such as water-assisted CVD~\cite{Hata-supergrowth,Noda-supergrowth-catalyst-support}, oxygen-assisted CVD~\cite{Zhang-Dai-roles-H2-O2}, point-arc microwave plasma CVD~\cite{Zhong-VASWNTs-plasmaCVD}, molecular-beam synthesis~\cite{Eres-molecular-beam-VASWNTs}, and hot-filament CVD~\cite{Xu-Hauge-hot_filament_VASWNTs}. It has also been shown that VA-SWCNTs can be produced using traditional thermal CVD methods if the catalyst concentration is optimized for the synthesis conditions~\cite{Noda-Co-Mo_combinatorial,Zhang-Resasco-catal_density}. These new approaches have succeeded in significantly increasing the overall yield, but improvements in SWCNT quality and control over chirality are still necessary, particularly when considering electrical and optical applications.

In a previous report~\cite{Maruyama-CPL403} we presented an {\em in situ\/} optical absorbance measurement technique, by which the thickness of a VA-SWCNT film could be measured during growth. In the present study, we employ this real-time technique to investigate the growth dynamics of VA-SWCNTs, and find that as growth progresses, the growth rate of the VA-SWCNTs exponentially decreases from an initial maximum. We have also systematically investigated the temperature and pressure effects on the initial film growth rate, and find the optimum growth pressure changes with the CVD temperature. Below this optimum pressure the growth reaction is first-order, with the arrival of ethanol at the catalyst being the rate-limiting step. These new findings are important in clarifying the processes governing the growth reaction. We also present a new method of measuring the burning temperature of a low-mass material, which combines the {\em in situ\/} absorbance measurement with controlled oxidation. This method is demonstrated using a portion of a VA-SWCNT film that has insufficient mass to measure using conventional thermogravimetric analysis.

\section{Experimental}
Films of VA-SWCNTs were synthesized on an optically polished quartz substrate that had been dip-coated~\cite{Murakami-CPL377} into Co and Mo acetate solutions (metal content 0.01 wt\,\% each). VA-SWCNTs were synthesized from ethanol at temperatures between 750 and 825\,\textcelsius\/, and the CVD time was typically 10-20 min. Details regarding the synthesis procedure have been reported elsewhere~\cite{Murakami-CPL385,Maruyama-CPL403,Murakami-PRL94,Murakami-Carbon43}. 

\begin{figure}[tb]
	\begin{center}
		\includegraphics[width=0.47\textwidth]{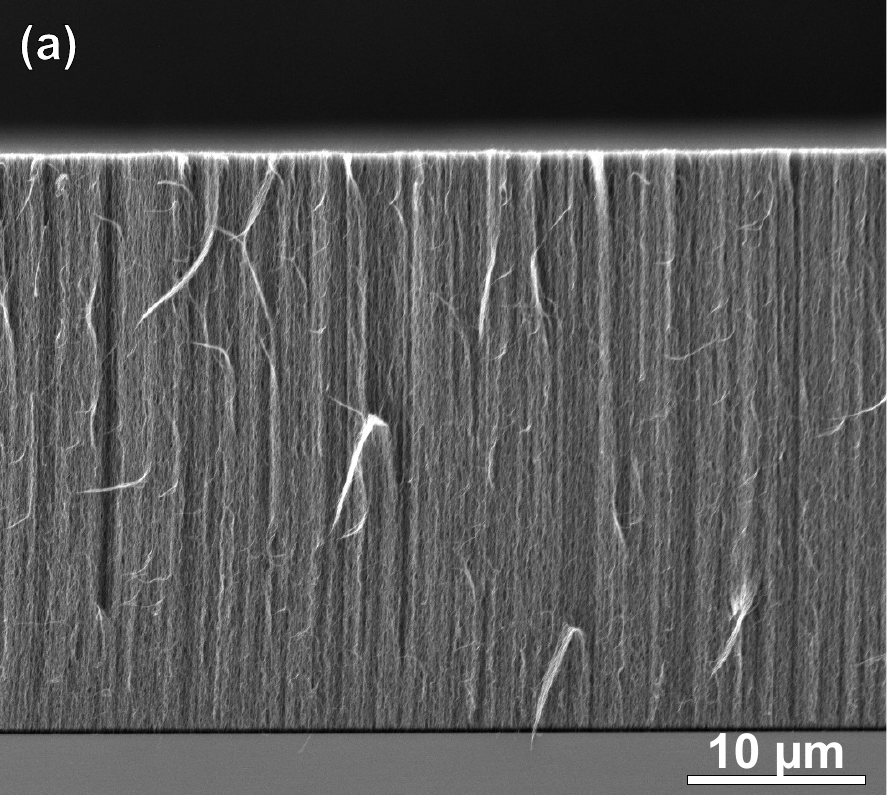}
		\quad
		\includegraphics[width=0.47\textwidth]{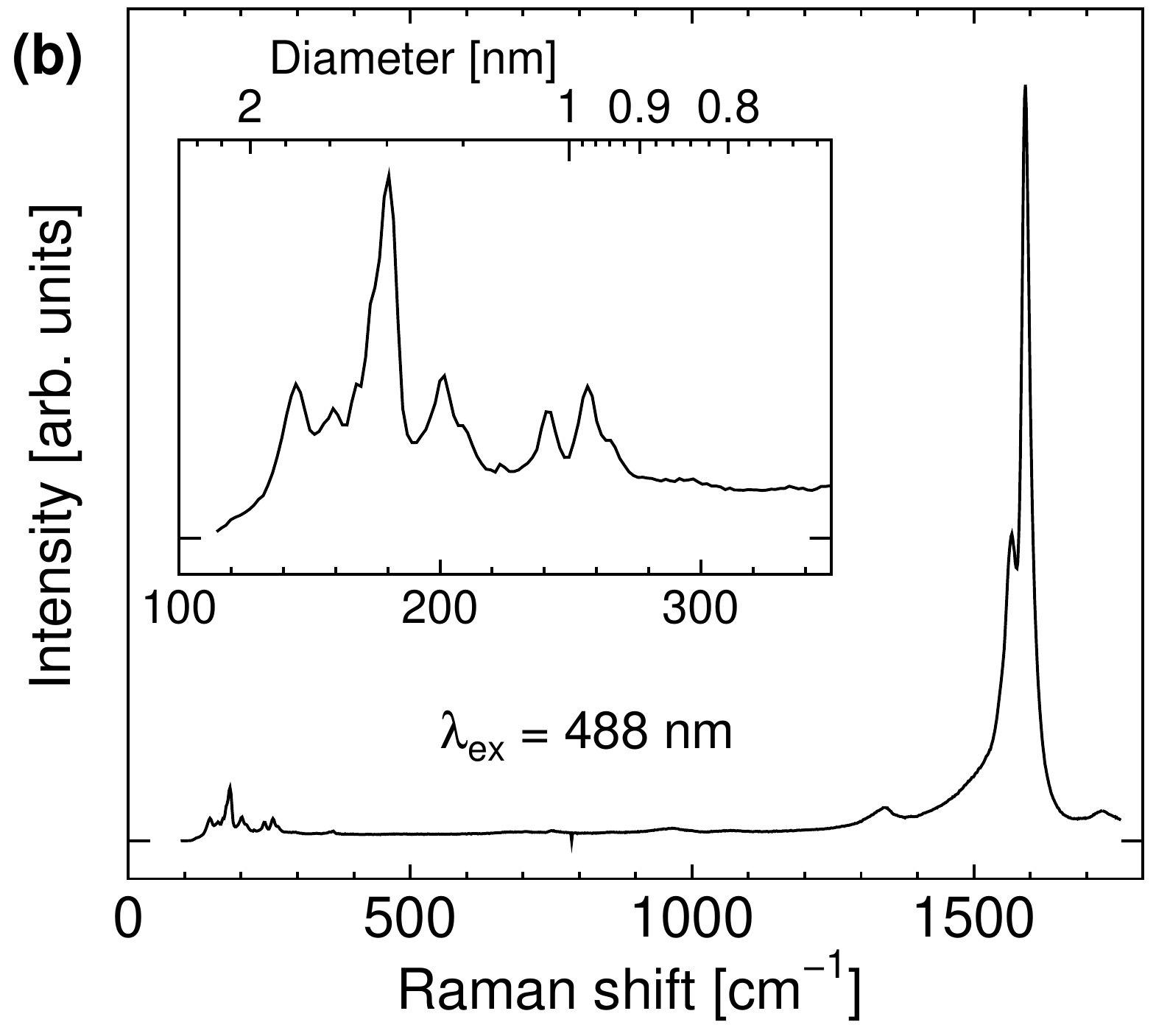}
		\caption{\label{SEM-Raman}(a) Cross-sectional SEM image of a VA-SWCNT film (thickness $\approx$\,32\,$\mu$m), and (b) a corresponding resonance Raman spectrum (excitation wavelength\,=\,488\,nm). The radial breathing mode peaks, with strong characteristic peak at 180\,cm$^{-1}$, are shown in the inset. \newline}
	\end{center}
\end{figure}

A cross-sectional scanning electron microscope (SEM) image of a VA-SWCNT film is shown in Fig.~\ref{SEM-Raman}a.
In the corresponding resonance Raman spectrum (Fig.~\ref{SEM-Raman}b, excitation wavelength $\lambda_{\mathrm{ex}}$\,=\,488\,nm), the presence of a strong peak at 1592\,cm$^{-1}$ (the G-band) and radial breathing mode (RBM) peaks between 100 and 350\,cm$^{-1}$ reveal the presence of SWCNTs (see e.g.~\cite{Dresselhaus-Eklund-Phonons}). The weak intensity of the D-band near 1340\,cm$^{-1}$ relative to the G-band reflects the high purity of the SWCNTs. In the RBM region, a dominant peak at 180\,cm$^{-1}$ indicates the SWCNTs are oriented normal to the substrate surface~\cite{Murakami-PRB71}. 

Growth of the VA-SWCNT film was recorded using the {\em in situ\/} optical method described in Ref.~\cite{Maruyama-CPL403}. Prior to growth, the dip-coated quartz substrate was positioned such that a laser (here, $\lambda$\,=\,488\,nm) was incident normal to the substrate through a small opening (dia.~$\approx$\,4\,mm) in the bottom of the CVD furnace. The intensity of the transmitted light was measured after passing through another small opening in the top of the furnace, and the absorbance of the VA-SWCNT film was determined using the Beer-Lambert law. 

\section{Results and Discussion}
\subsection{Description of the growth process}

In the alcohol CVD process, synthesis of VA-SWCNTs occurs by a root-growth mechanism~\cite{XiangRong-13C_root_growth}, where ethanol molecules react with metal catalyst nanoparticles on the substrate surface. The amount of carbon supplied to the catalyst is the flux, $J$ [mol\,$\mu$m$^{-2}$\,s$^{-1}$], at the substrate surface. Through the catalytic reaction, $M$~[mol\,$\mu$m$^{-2}$\,s$^{-1}$] moles of this available carbon are converted into SWCNTs. This outflux of carbon in the form of SWCNTs is the molar growth rate of the VA-SWCNT film per substrate area.

During growth, the thickness of the VA-SWCNT film is determined from its optical absorbance, $A$. The absorbance of the film can be expressed as $A\,=\,\varepsilon\rho\ell$, where $\varepsilon$~[$\mu$m$^{2}$\,mol$^{-1}$] is the molar absorption cross-section, $\rho$~[mol\,$\mu$m$^{-3}$] is the molar density of carbon in the film, and $\ell$~[$\mu$m] is the optical path length through the absorbing material, i.e.~the total film thickness. Since the dip-coat method produces monodispersed catalyst nanoparticles on the substrate surface~\cite{Hu-J_Catal}, the SWCNT concentration within the area of the laser spot is assumed uniform. The product $\varepsilon\rho$ is the absorption coefficient $\alpha$~[$\mu$m$^{-1}$], and is a function of wavelength. Experimentally, we have found $\alpha(\lambda\!=\!488\,$nm) $\approx\,0.147\,\mu$m$^{-1}$. The thickness of the VA-SWCNT film {\em per side\/} of the substrate, $L$ [$\mu$m], is defined by $2L\,\equiv\,\ell$, which yields the more practical relation $L\,\approx\,A\,\times\,6.78\,\mu$m. 

\begin{figure}[b]
	\begin{center}
		\includegraphics[width=0.5\textwidth]{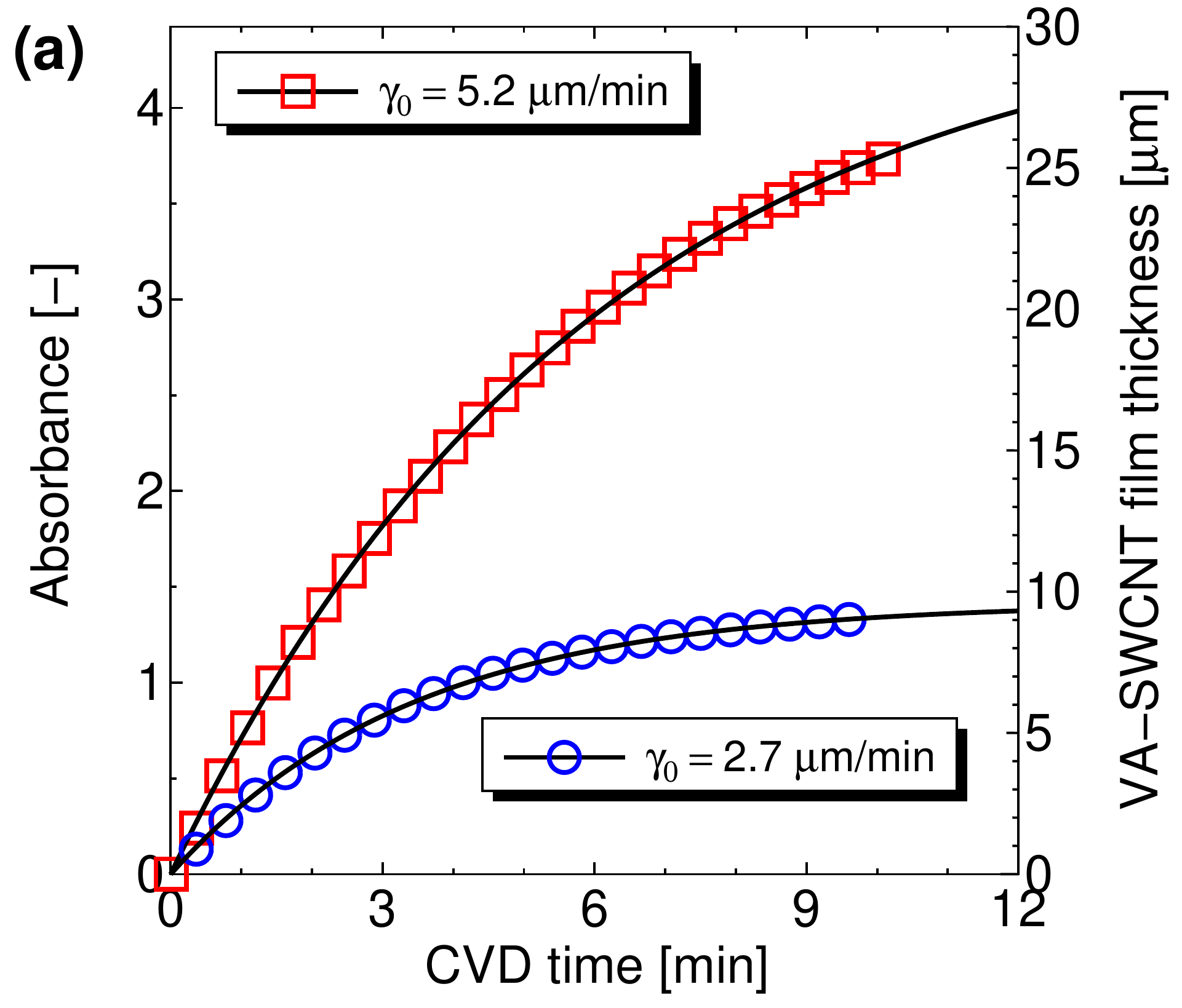}
		\quad
		\includegraphics[width=0.46\textwidth]{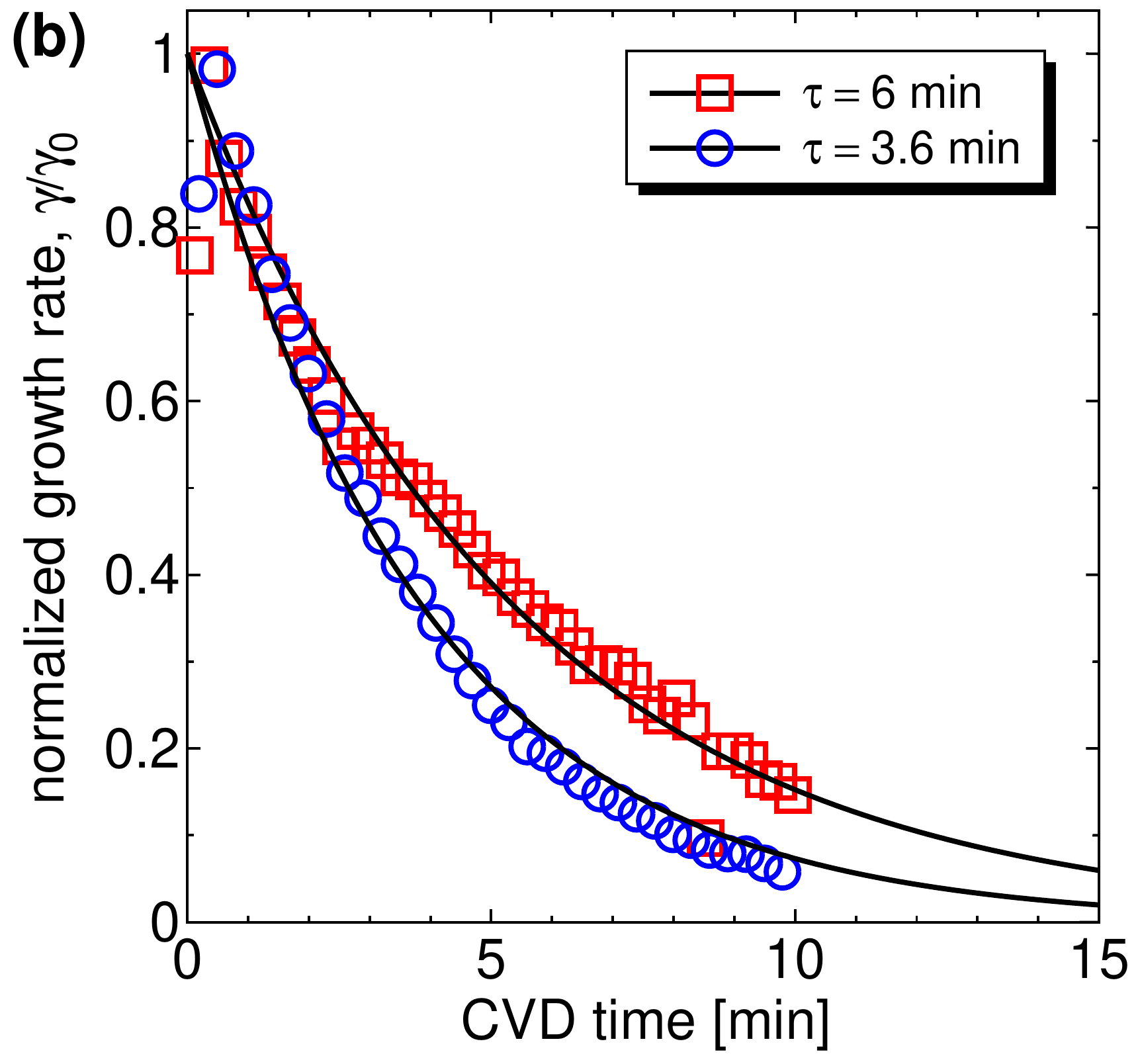}
		\caption{\label{abs-fit} (a) Optical absorbance and corresponding thickness of VA-SWCNT films during CVD. Fitting curves correspond to Eq.~(\ref{film-thickness}). (b) Normalized growth rates for the two cases shown in (a). Corresponding curves were calculated using values of $\tau$ determined from fitting.}
	\end{center}
\end{figure}

Optical absorbance data for two different VA-SWCNT films measured during CVD synthesis are plotted in Fig.~\ref{abs-fit}a.
The right ordinate shows the thickness of the film, which was determined by scaling the absorbance (left ordinate) by 6.78 $\mu$m. The result is a growth profile of the VA-SWCNT film. The growth rate of the film, $\gamma$ [$\mu$m\,s$^{-1}$], is the time-derivative of the thickness, $\gamma(t)\,=\,\frac{dL(t)}{dt}$. Normalized growth rates, obtained by differentiating the absorbance data, are plotted in Fig.~\ref{abs-fit}b. It is clear that as the reaction progresses, the growth rate decreases exponentially from an initial maximum. One possible explanation for this behavior is that the growing VA-SWCNT film acts as a diffusion barrier, impeding the supply of ethanol to the catalyst. For VA-SWCNT films of the thickness discussed in this report, however, the diffusion resistance imposed by the film is negligible~\cite{XiangRong-diffusion_limit}. The more plausible explanation for the exponential decay of the growth rate is some form of catalyst poisoning, which may be due to the formation of a carbonaceous layer around the catalyst nanoparticles, or secondary reactions with byproducts of ethanol dissociation. In this case, the rates of these catalyst poisoning processes are proportional to the rate of the synthesis reaction. Hence, we postulate that the SWCNT synthesis reaction is self-exhausting, expressed by
\begin{equation}
	\frac{d\gamma}{dt} = -\frac{1}{\tau} \gamma,
	\label{catal-act-decay}
\end{equation}
where $\tau$~[s] is the reaction time constant. Solving (\ref{catal-act-decay}) yields the time-dependent expression for the growth rate, 
\begin{equation}
	\gamma(t) = \gamma_{0}\exp\Bigl(\frac{-t}{\tau}\Bigr),
	 \label{film-growth-rate}
\end{equation}
where $\gamma_{0} = \gamma(t\!=\!0)$ is the initial growth rate of the VA-SWCNT film. This is the typical form of a decay process, and is expected based on our assumption of irreversible catalyst poisoning (i.e.~no reactivation mechanism). The overall thickness of the synthesized VA-SWCNT film is the integral of the growth rate. Taking into account the initial condition $L(t\!=\!0)\,=\,0$, the overall thickness $L(t)\,=\,\int{\gamma(t)dt}$ is
\begin{equation}
	L(t) = \gamma_{0}\tau \biggl[ 1-\exp \Bigl( \frac{-t}{\tau} \Bigr) \biggr].
	\label{film-thickness}
\end{equation}
This equation shows the maximum obtainable film thickness is $L(t\!\rightarrow\!\infty)\,=\,\gamma_{0}\tau$, which is appropriately determined by the two fitting parameters.

The data in Fig.~\ref{abs-fit}a were fitted using Eq.~(\ref{film-thickness}). The growth rates for these two cases are plotted in Fig.~\ref{abs-fit}b. The superimposed lines are calculations using Eq.~(\ref{film-growth-rate}) and values of $\tau$ determined from fitting. The results of many such fittings reveal a wide range of initial growth rates, spanning an order-of-magnitude, but much less variation in the time constant, which typically has values of $\tau \approx\,4.5\,\pm\,2.5$ min. For these values, the effective growth time is approximately 15 min (see Fig.~\ref{catal-act-decay}b).

Equation (\ref{film-thickness}) has the same form as the simplified kinetic model described by Puretzky {\em et al.\/}~\cite{Puretzky-kinetic_model}, which disregards secondary catalyst deactivation processes. Based on this model,
\begin{equation}
	\tau \propto \frac{k_{sb}}{k_{cl}}\frac{1}{J},
		\label{tau-meaning}
\end{equation}
where $k_{sb}$ and $k_{cl}$ are the rates of incorporation of carbon into the catalyst nanoparticle and formation of a carbonaceous layer on the catalyst surface, respectively, and $J$ is the incoming flux of carbon to the catalyst.
The initial growth rate $\gamma_{0}$ is described by
\begin{equation}
	\gamma_{0} \propto J\exp\Bigl(\frac{-E_{a}}{k_{\mathrm{B}}T}\Bigr),
		\label{gamma-meaning}
\end{equation}
revealing that $\gamma_{0}$ is proportional to the flux of carbon at the catalyst surface. 

A different perspective on the growth can be obtained by replacing $\gamma$ with $\frac{dL}{dt}$ and then integrating Eq.~(\ref{catal-act-decay}). This yields
\begin{equation}
	\gamma(t) = \gamma_{0} - \frac{L(t)}{\tau},
		\label{growth-rate-vs-thickness}
\end{equation}
which indicates the growth rate decreases linearly with increasing film thickness. 
This suggests the diminishing growth rate is caused by reaction-driven catalytic deactivation rather than diffusion-limited transport. Without catalytic deactivation, the growth rate is expected to be constant in the case of no diffusion resistance, and proportional to $t^{-1/2}$ (where $t$ is the CVD time) in the strong diffusion-limited regime~\cite{XiangRong-diffusion_limit}. The latter likely applies to mm-thick VA-SWCNT films such as the so-called `supergrowth'~\cite{Hata-supergrowth}, even though the dependence of the final film thickness on CVD time was interpreted as catalytic poisoning~\cite{Futaba-supergrowth_kinetics}. For the purpose of controlling and optimizing VA-SWCNT synthesis it is important to discriminate between these behaviors, which have different origins and impose different limitations.

It is interesting to note that the time constant $\tau$ obtained from the alcohol catalytic CVD method ($\sim$\,4.5 min) happens to be similar to that reported by Futaba {\em et al.\/}~\cite{Futaba-supergrowth_kinetics} for catalytic CVD of ethylene. Other CVD methods, however, report nearly linear growth rates~\cite{Zhang-Dai-roles-H2-O2,Zhong-VASWNTs-plasmaCVD}, thus a $\tau$ of several hours. This suggests the reaction mechanism and effective growth time are dependent on the synthesis method.
One possible explanation for the slow growth rate found for alcohol CVD is that pure ethanol does not supply the proper balance of carbon, hydrogen, and oxygen. A number of groups have reported~\cite{Hata-supergrowth,Noda-supergrowth-catalyst-support,Zhang-Dai-roles-H2-O2,Xu-Hauge-hot_filament_VASWNTs} that the proper carbon-hydrogen-oxygen balance is critical to the synthesis of mm-long VA-SWCNT films, so the addition of one or more of these elements may significantly increase the growth rate and yield of VA-SWCNTs synthesized by alcohol CVD.

\subsection{Effect of synthesis conditions on growth parameters}
In this section we present the results of a series of CVD experiments systematically carried out to investigate the influence of the growth environment on the initial growth rate $\gamma_{0}$ and the reaction time constant $\tau$. 

\begin{figure}[t]
	\begin{center}
		\includegraphics[width=0.47\textwidth]{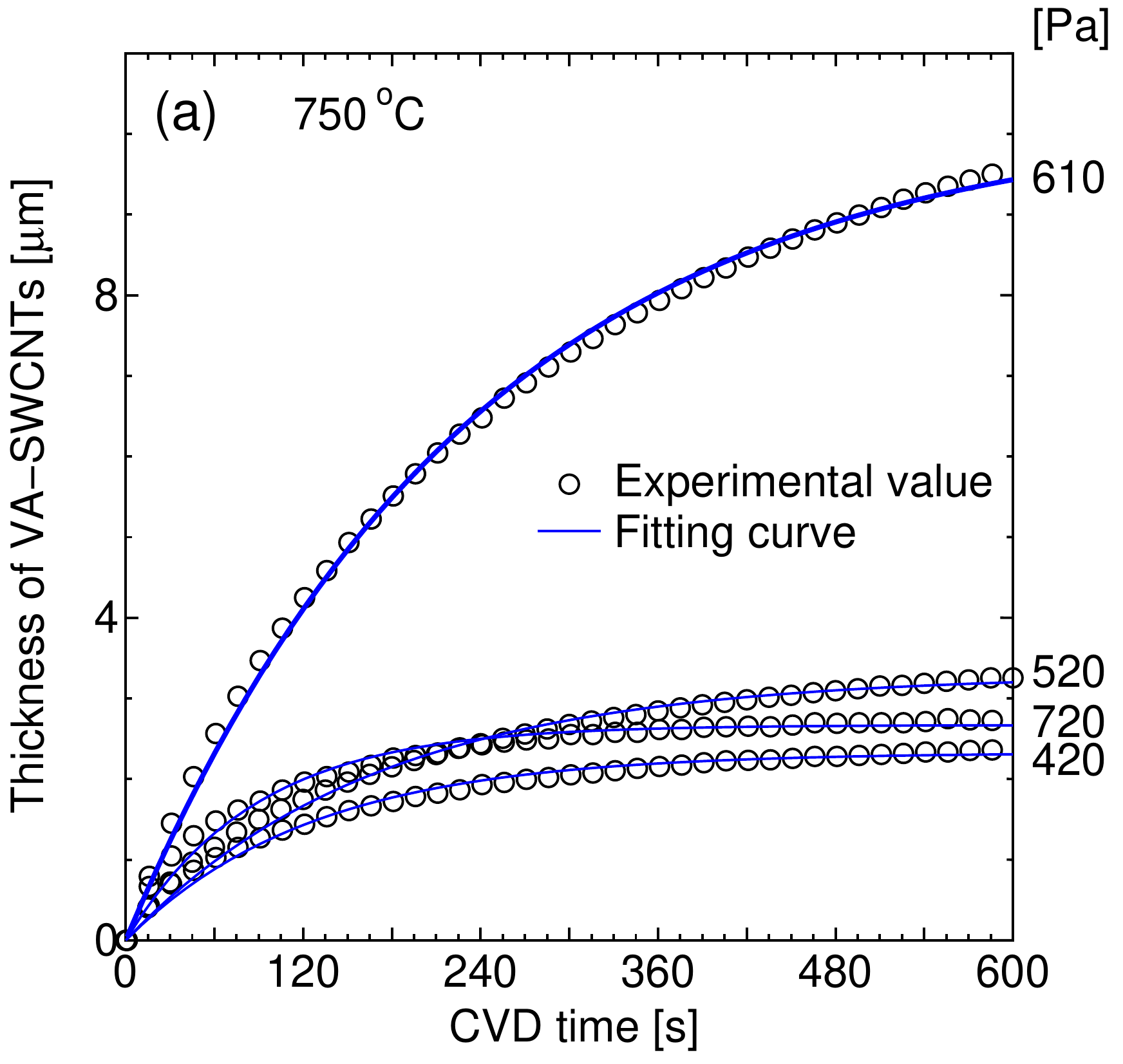}
		\quad
		\includegraphics[width=0.47\textwidth]{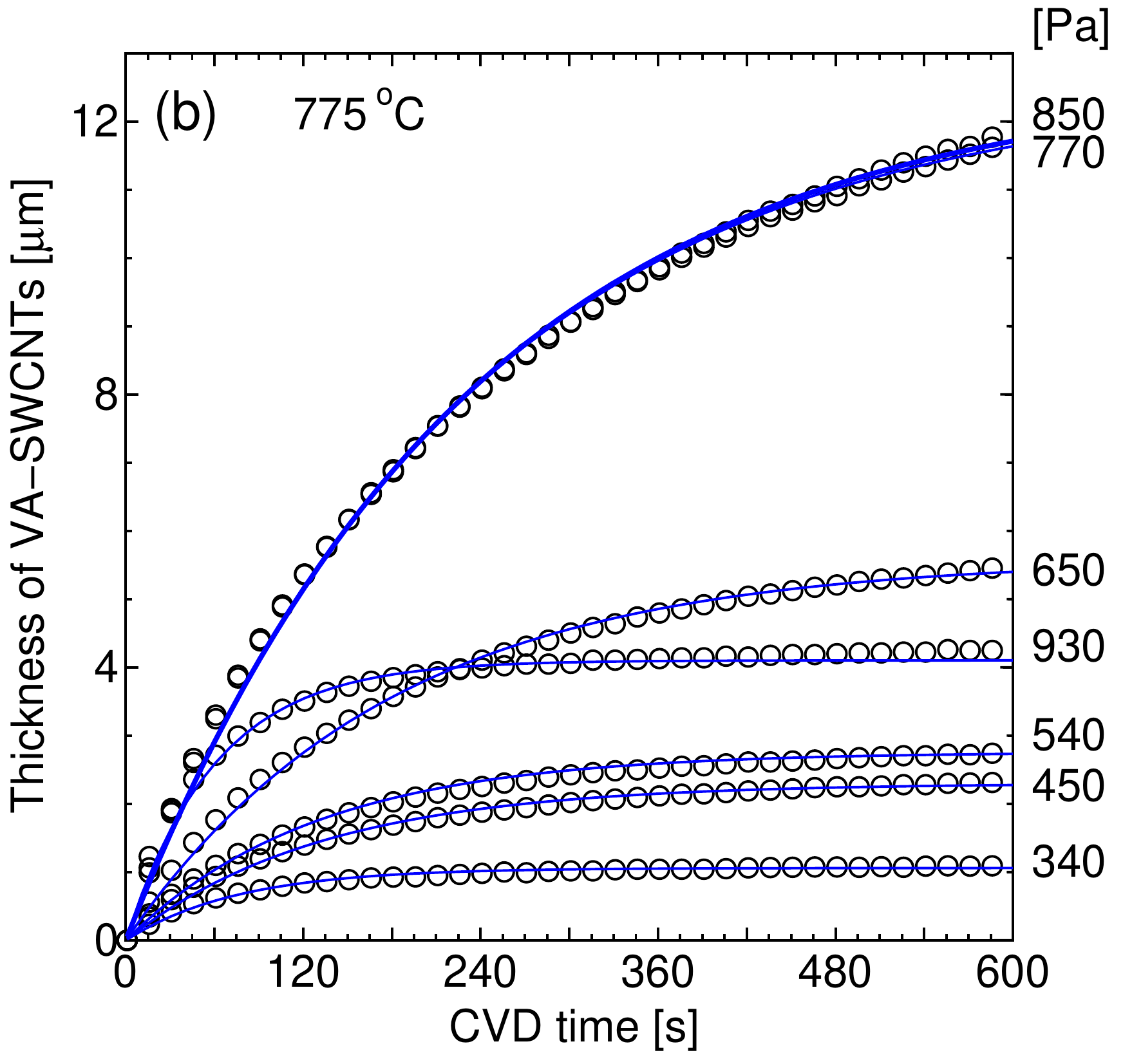}
		\\
		\includegraphics[width=0.47\textwidth]{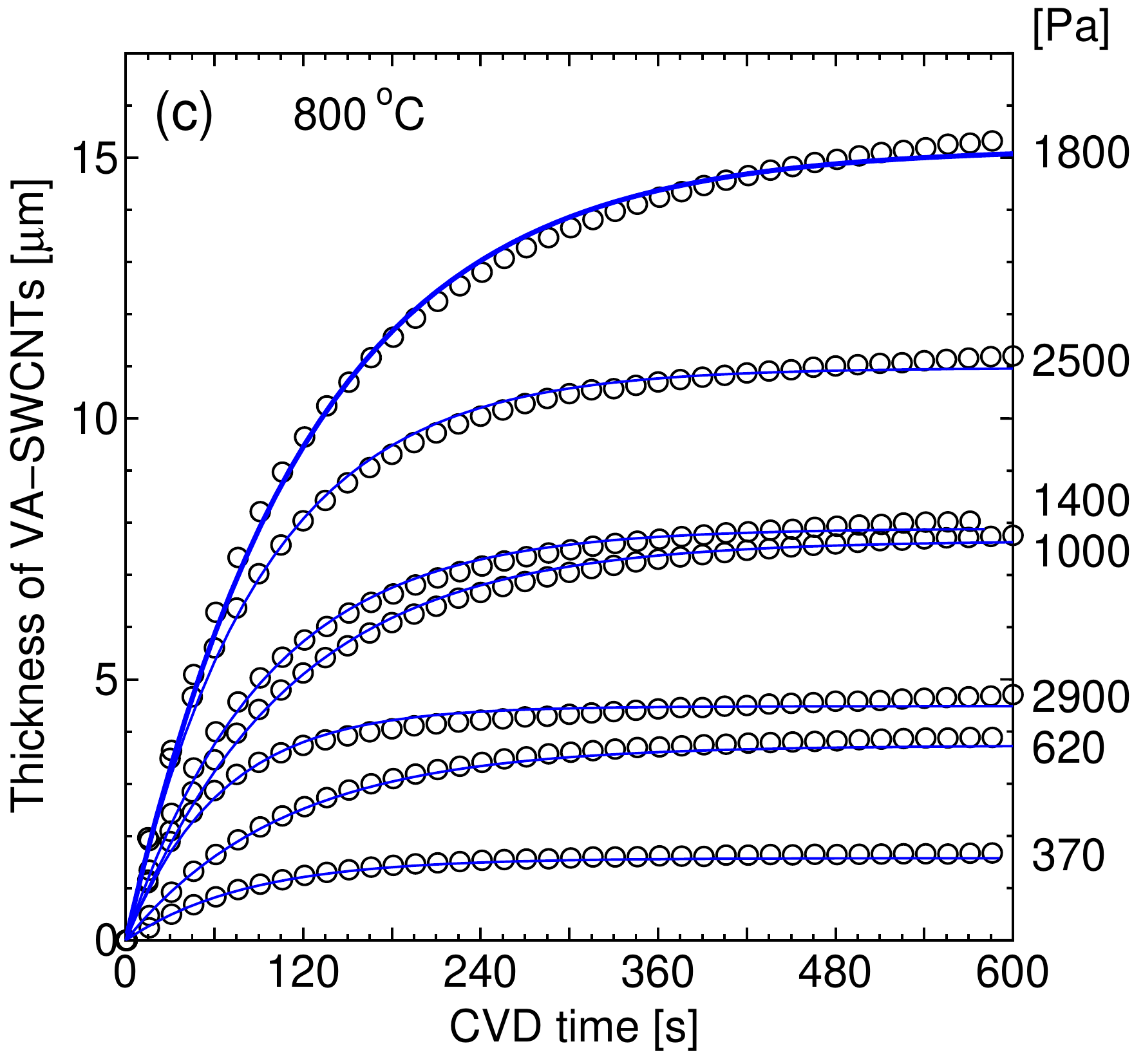}
		\quad
		\includegraphics[width=0.47\textwidth]{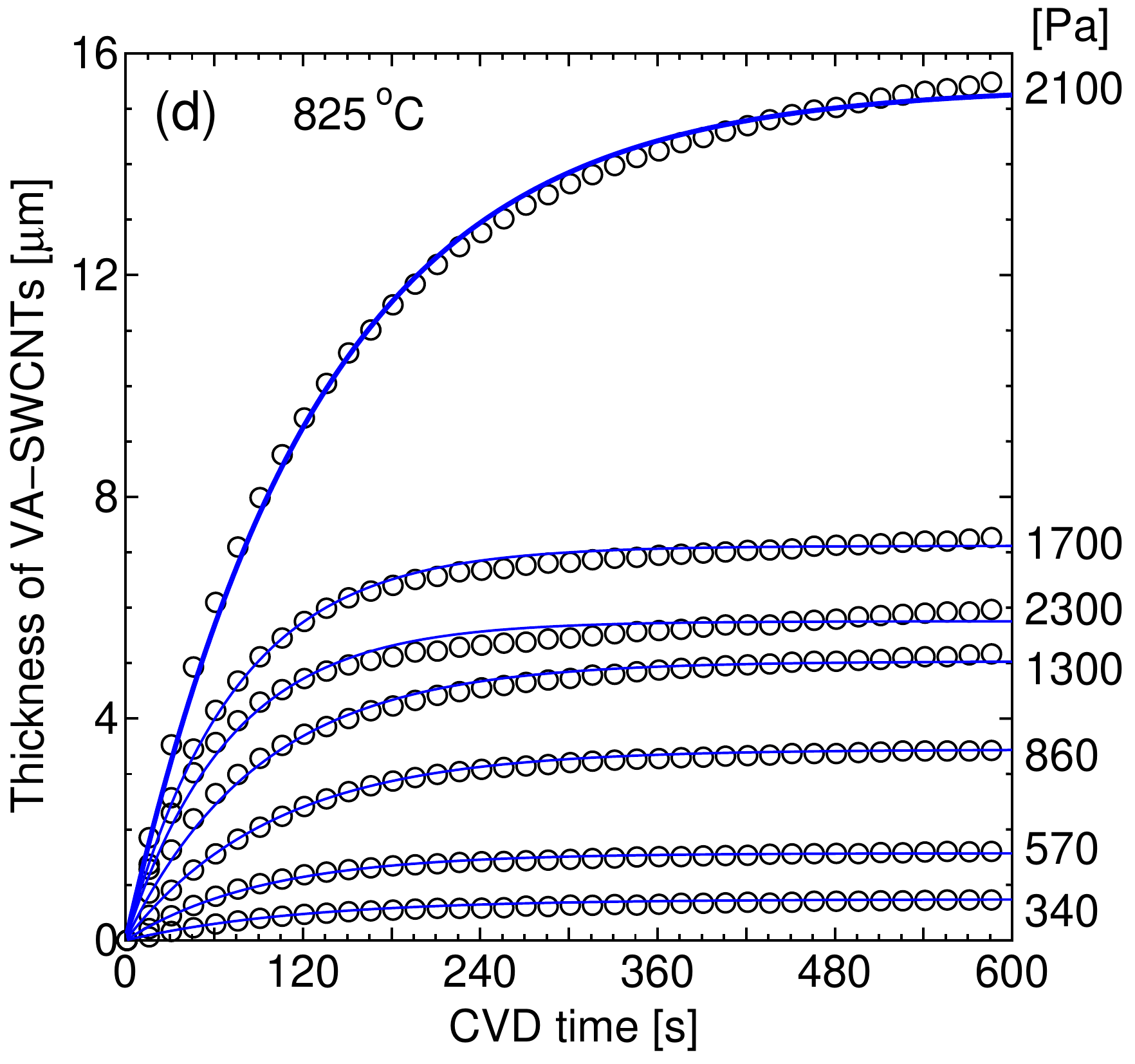}
		\caption{\label{pressure-growth} Growth profiles corresponding to VA-SWCNT synthesis at (a)~750, (b)~775, (c)~800, and (d)~825\,\textcelsius\/ for a range of ethanol pressures (values listed at right).}
	\end{center}
\end{figure}

\begin{figure}[b]
	\begin{center}
		\includegraphics[width=0.47\textwidth]{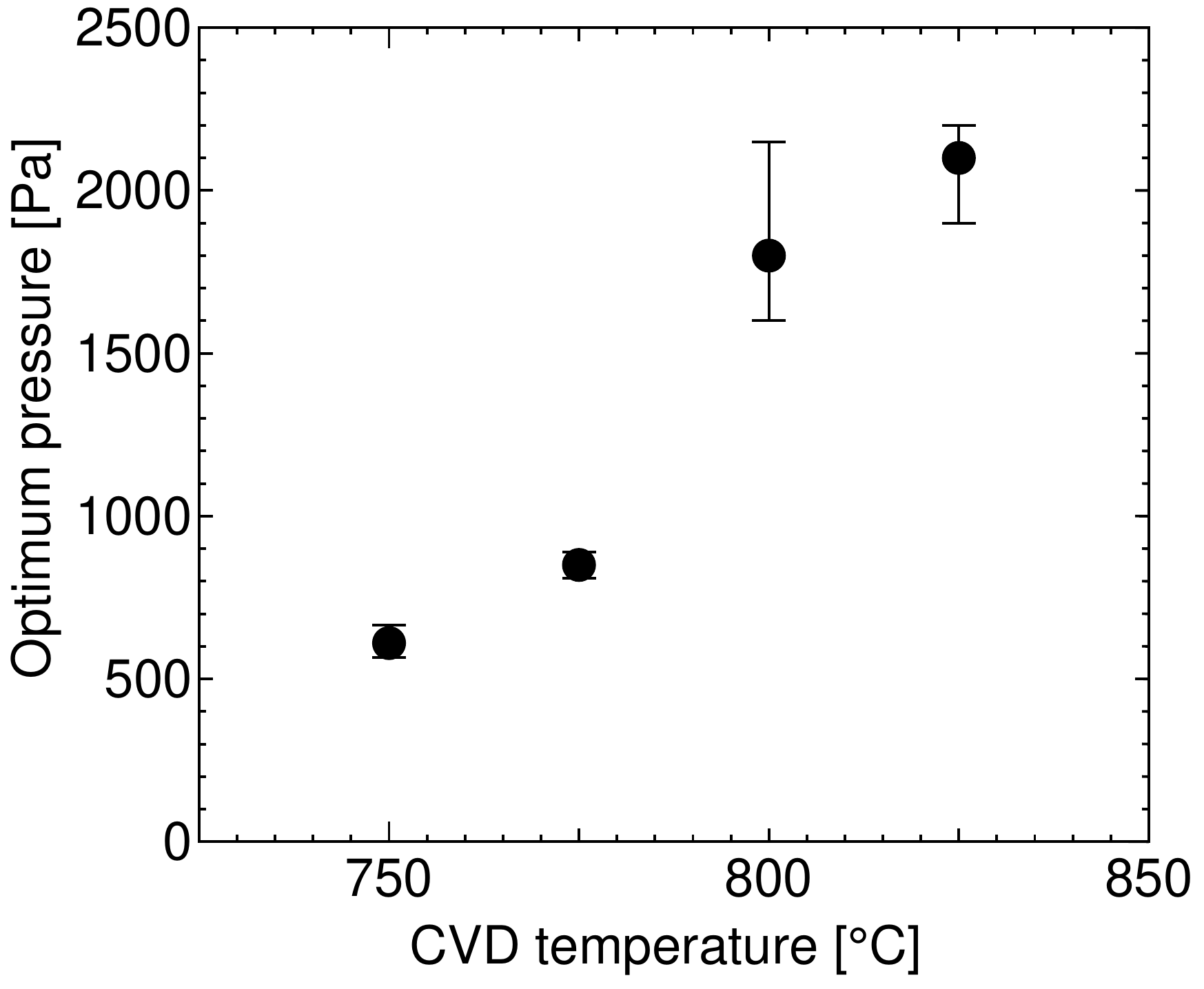}
		\caption{\label{opt}Optimum ethanol pressure $P_{\mathrm{opt}}$ for different CVD temperatures.}
	\end{center}
\end{figure}

A series of growth profiles corresponding to different CVD temperatures and ethanol pressures are shown in Figs.~\ref{pressure-growth}a -- \ref{pressure-growth}d.
In all cases, the ethanol flow rate was 500\,sccm. The circles represent experimental values, while the curves were fitted using Eq.~(\ref{film-thickness}). The main result in Fig.~\ref{pressure-growth} is that for each growth temperature there is a different optimum pressure, $P_{\mathrm{opt}}$, at which overall VA-SWCNT growth is maximized. Furthermore, $P_{\mathrm{opt}}$ increases with CVD temperature, as shown in Fig.~\ref{opt}. 

\begin{figure}[tb]
	\begin{center}
		\includegraphics[width=0.5\textwidth]{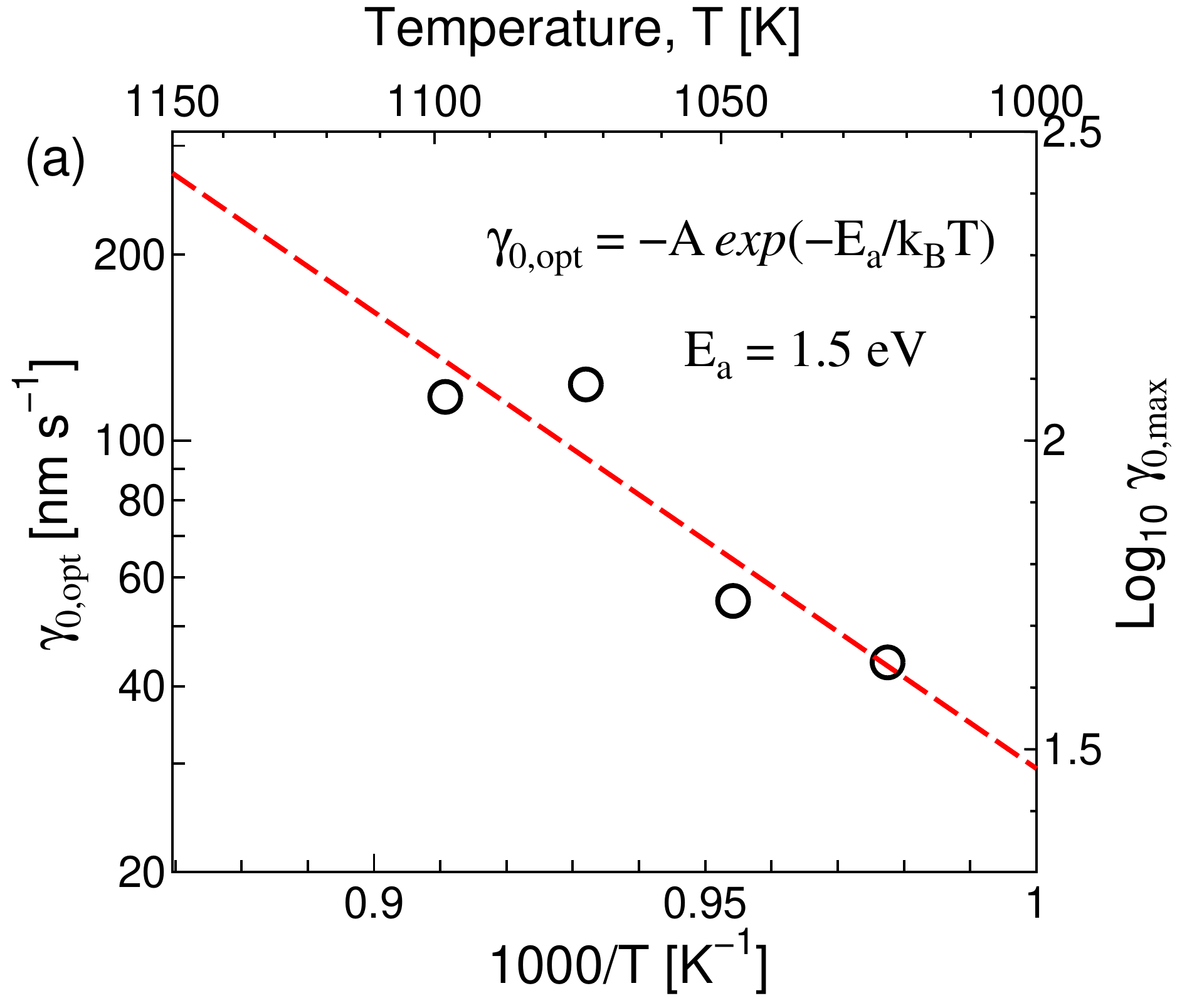}
		\quad
		\includegraphics[width=0.45\textwidth]{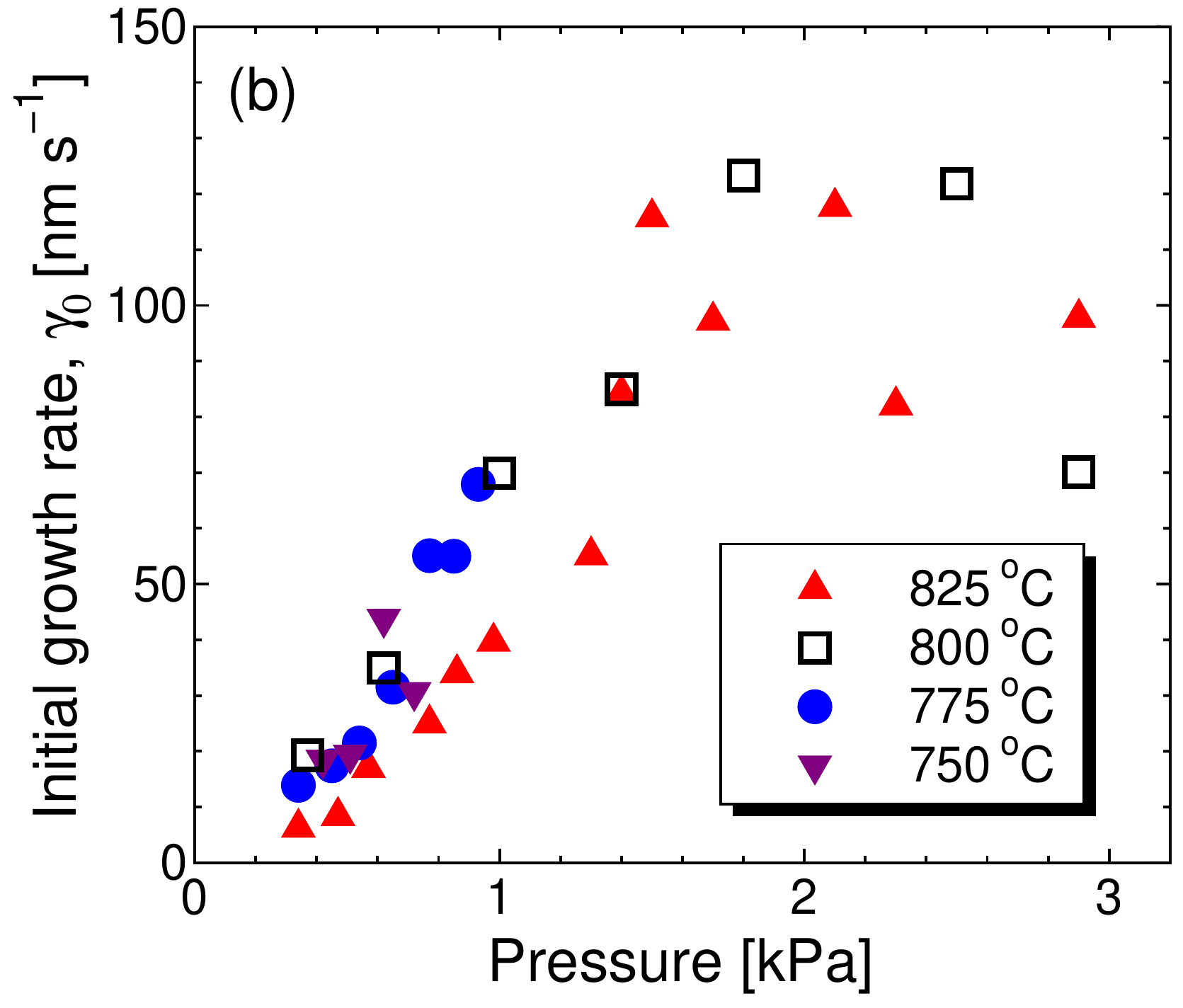}
		\caption{\label{Arrhenius}(a) Arrhenius plot based on the values of $\gamma_{0}$ determined for the optimum growth pressures, $P_{\mathrm{opt}}$. The activation energy, $E_{a}$ is $\sim$1.5\,eV. (b)~Initial growth rate as a function of ethanol pressure for different CVD temperatures. Below $P_{\mathrm{opt}}$, the growth rate is proportional to pressure, typical of a first-order reaction. \newline}
	\end{center}
\end{figure}

Equation (\ref{gamma-meaning}), which describes the film growth rate, is an Arrhenius equation, with the prefactor proportional to the flux, and an overall activation energy $E_{a}$. An Arrhenius plot of $\gamma_{0}$ is shown in Fig.~\ref{Arrhenius}.
The data are the initial growth rates corresponding to $P_{\mathrm{opt}}$ for each temperature. The data do not fall exactly on a line, but follow a linear trend. The activation energy determined from this linear approximation is 1.5\,eV, which is slightly lower than the overall activation energy of 2\,eV reported by Puretzky {\em et al.\/}~\cite{Puretzky-kinetic_model}. We can speculate that this value is directly comparable to the 1.6\,eV activation energy for the bulk diffusion of carbon through a metal catalyst nanoparticle to the growth edge of a nanotube~\cite{Puretzky-kinetic_model,Louchev-CVD-diffusion-kinetics}. Even in the case of surface diffusion, the energy required to be incorporated into the graphitic structure of the SWCNT should be comparable.

Figure \ref{Arrhenius}b shows values of the initial growth rate $\gamma_{0}$ as a function of ethanol pressure for different CVD temperatures. The behavior is more apparent at 800 and 850\,\textcelsius, but for a given CVD temperature, $\gamma_{0}$ is proportional to pressure up to $P_{\mathrm{opt}}$ ($\sim$\,1.8\,kPa for 800\,\textdegree\/). This indicates that below $P_{\mathrm{opt}}$ the synthesis reaction is limited by the rate at which ethanol molecules arrive at active sites on the catalyst. This means the growth of VA-SWCNTs in this pressure regime is governed by a first-order reaction. Above $P_{\mathrm{opt}}$ the behavior changes, presumably because the flux of ethanol at the catalyst is more than sufficient to sustain the reaction, so a different step in the overall process becomes the rate-limiting step.

\subsection{Burning and the growth environment}
Although VA-SWCNT growth usually occurs as described in the preceding discussion, where the film approaches its maximum thickness as $t\rightarrow\infty$, an apparent {\em decrease\/} in film thickness has been observed~\cite{Maruyama-CPL403} for CVD times of 30 min or longer. It was hypothesized that this decrease may be due to burning of the VA-SWCNT film, possibly caused by a slow leak of air into the CVD chamber. However, since $\gamma_{0}$ and $\tau$ can vary for VA-SWCNTs synthesized under similar conditions, it is also possible that one or both of these parameters were slightly lower at the onset of growth for those films grown for longer CVD times, thus reaching a shorter maximum height and giving the illusion of a decrease in film thickness. In order to clarify whether or not burning is occurring, we monitored the growth process of VA-SWCNTs synthesized under different vacuum conditions. Immediately prior to heating the CVD furnace, the room-temperature leak rate into the CVD chamber was measured. The chamber was then heated to 800\,\textcelsius\/, and ethanol was introduced. When the growth rate had diminished to essentially zero, the ethanol flow was stopped and the CVD chamber was held at 800\,\textcelsius\/. 

\begin{figure}[b]
	\begin{center}
		\includegraphics[width=0.47\textwidth]{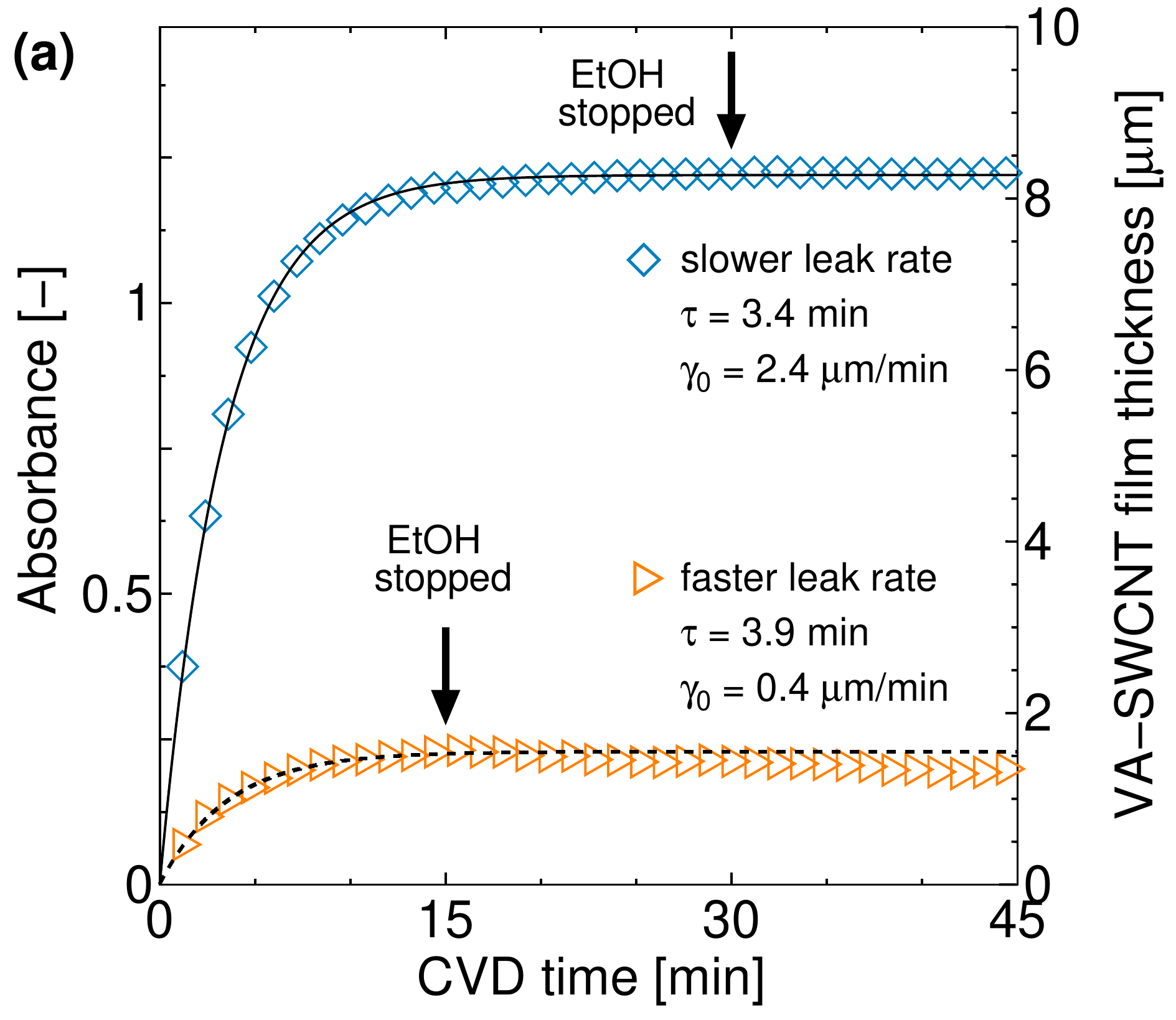}
		\quad
		\includegraphics[width=0.47\textwidth]{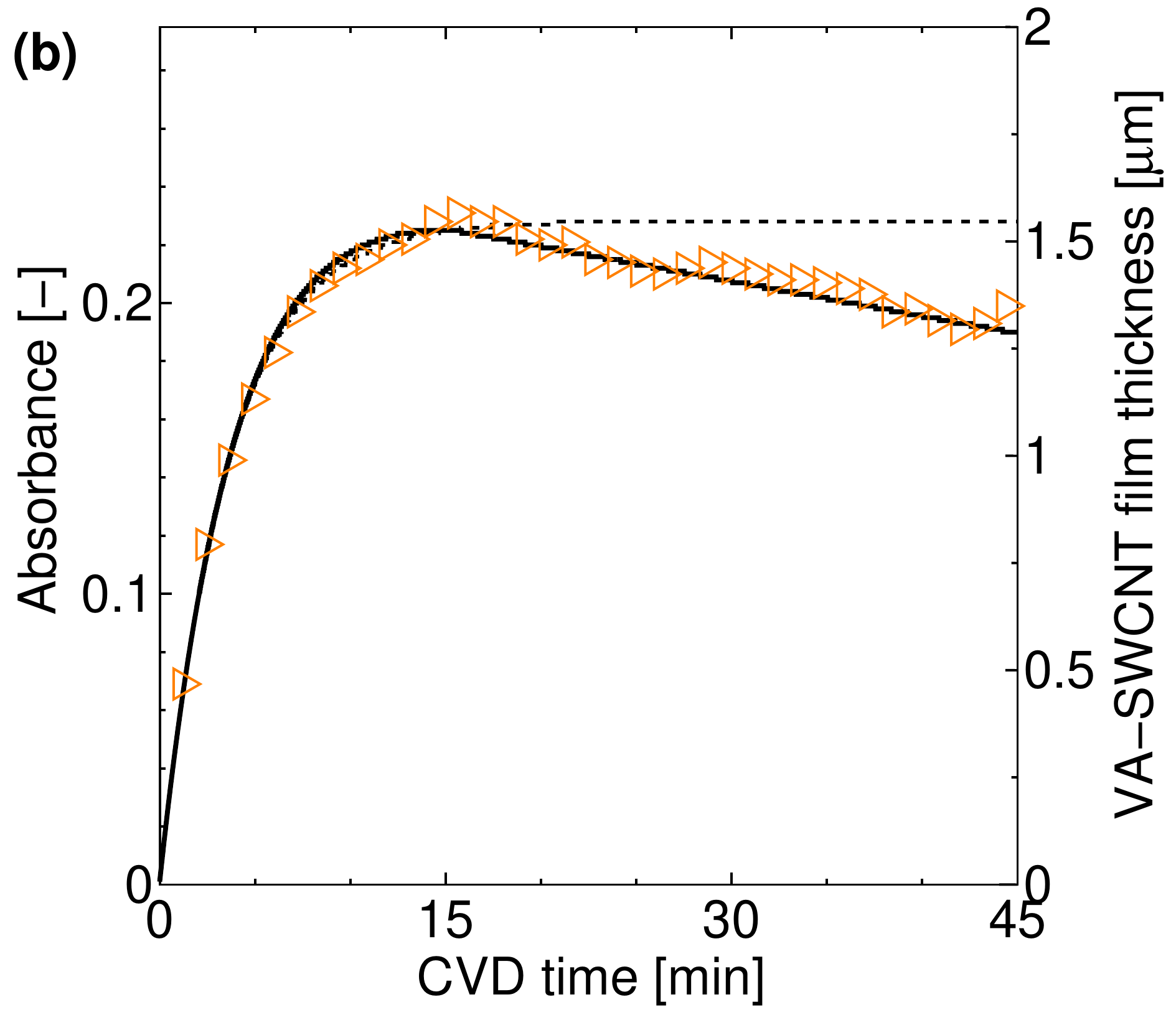}
		\caption{\label{fit-burning}(a) Two cases where the leak rate of air into the CVD chamber was slow (upper series, diamonds), and fast (lower series, triangles). In both cases, the ethanol supply was cut-off after growth had effectively stopped (indicated by the arrows) and the CVD chamber was held at 800\,\textcelsius\/. A decrease in absorbance detected for the fast leak case (triangles) is shown in detail in (b), where the dashed curve was fitted using Eq.~(\ref{film-thickness}), and the solid curve using Eq.~(\ref{withbeta}), which accounts for burning.}
	\end{center}
\end{figure}

Growth profiles corresponding to two different leak rates are shown in Fig.~\ref{fit-burning}a, with an arrow indicating where the ethanol flow was stopped in each case. 
The upper series (diamonds) corresponds to a leak rate of 2.6x10$^{-3}$\,sccm. For this case, there was no measurable change in absorbance after stopping the ethanol flow. The lower series (triangles) corresponds to a case where the leak rate was more than three times faster (8.4x10$^{-3}$\,sccm). After cutting off the ethanol supply at $t$\,=\,15\,min, a decrease in the absorbance was soon apparent. This decrease is shown in detail in Fig.~\ref{fit-burning}b. We find $\tau$ is similar for both cases, but $\gamma_{0}$ is more than five times larger for the slow-leak case (diamonds). This indicates the initial growth rate is sensitive to the CVD environment immediately prior to growth. 

Burning of the VA-SWCNTs is evidenced by the decrease in optical absorbance, but it is not known where this burning occurs. It is possible that the top of the VA-SWCNT film is burned preferentially, causing the film thickness to decrease, but defective or small-diameter SWCNTs may burn at lower temperatures, leading to an overall `thinning' of the entire film. Another possibility is that the presence of oxygen at the catalyst may initiate burning at the base of the film. This issue has been addressed in a separate study~\cite{Hai-Thermal_degradation}, so for simplicity in analysis, we treat burning as if it were causing a decrease in the thickness of the film but no change in its density. Making this assumption, the discussion in the previous section can be extended to include burning of the SWCNTs simply by subtracting a constant term $\beta$ [$\mu$m\,s$^{-1}$] from the growth term in Eq.~(\ref{film-growth-rate}). This yields
\begin{equation}
	\gamma(t) = \gamma_{0}\exp\Bigl(\frac{-t}{\tau}\Bigr)-\beta,
	\label{film-growth-rate-beta}
\end{equation}
where $\beta$ effectively opposes the growth rate $\gamma(t)$. The resulting expression for the film thickness including burning effects is
\begin{equation}
	L(t) = \gamma_{0}\tau\biggl[ 1-\exp\Bigl(\frac{-t}{\tau}\Bigr) \biggr]-\beta t.
	\label{withbeta}
\end{equation}
Since burning obviously cannot occur if there are no SWCNTs, the conditions imposed on $\beta$ are 
\begin{equation}
	\beta \; 
	\begin{cases}
		\; = 0 &\text{when\;} L(t) = 0\\
		\; \geq 0 &\text{when\;} L(t)\,>\,0.
	\end{cases}
\end{equation}
The physical interpretation of $\beta$ is oxidation of the SWCNTs. Since the conditions inside the CVD chamber are unchanged during growth, it is reasonable to assume the burning rate is constant. The linear trend observed in Fig.~\ref{fit-burning}b also suggests this is the case. 

The data in Fig.~\ref{fit-burning}a were fit using both Eq.~(\ref{film-thickness}), in which burning effects are ignored ($\beta$\,=\,0), and Eq.~(\ref{withbeta}), which accounts for burning ($\beta\!>\!0$). For the slower leak case (diamonds), both fitting methods yield the same result, with $\beta\!\approxeq\!0$ in Eq.~(\ref{withbeta}). Similarly, fittings for the faster leak case are shown in detail in Fig.~\ref{fit-burning}b. When burning is ignored (dashed line), the early growth is well-described, but after stopping the ethanol flow, the data are only fit accurately if burning effects are included (solid line). The value of $\beta$ obtained from this fitting is approximately 2\% of $\gamma_{0}$ ($\beta\,\approx\,0.008\,\mu$m/min). This value is quite small, meaning burning would not be noticeable for short CVD times, but can become significant after long CVD times when the growth rate has diminished to essentially zero and only burning remains.

\begin{figure}[tb]
	\begin{center}
		\includegraphics[width=0.48\textwidth]{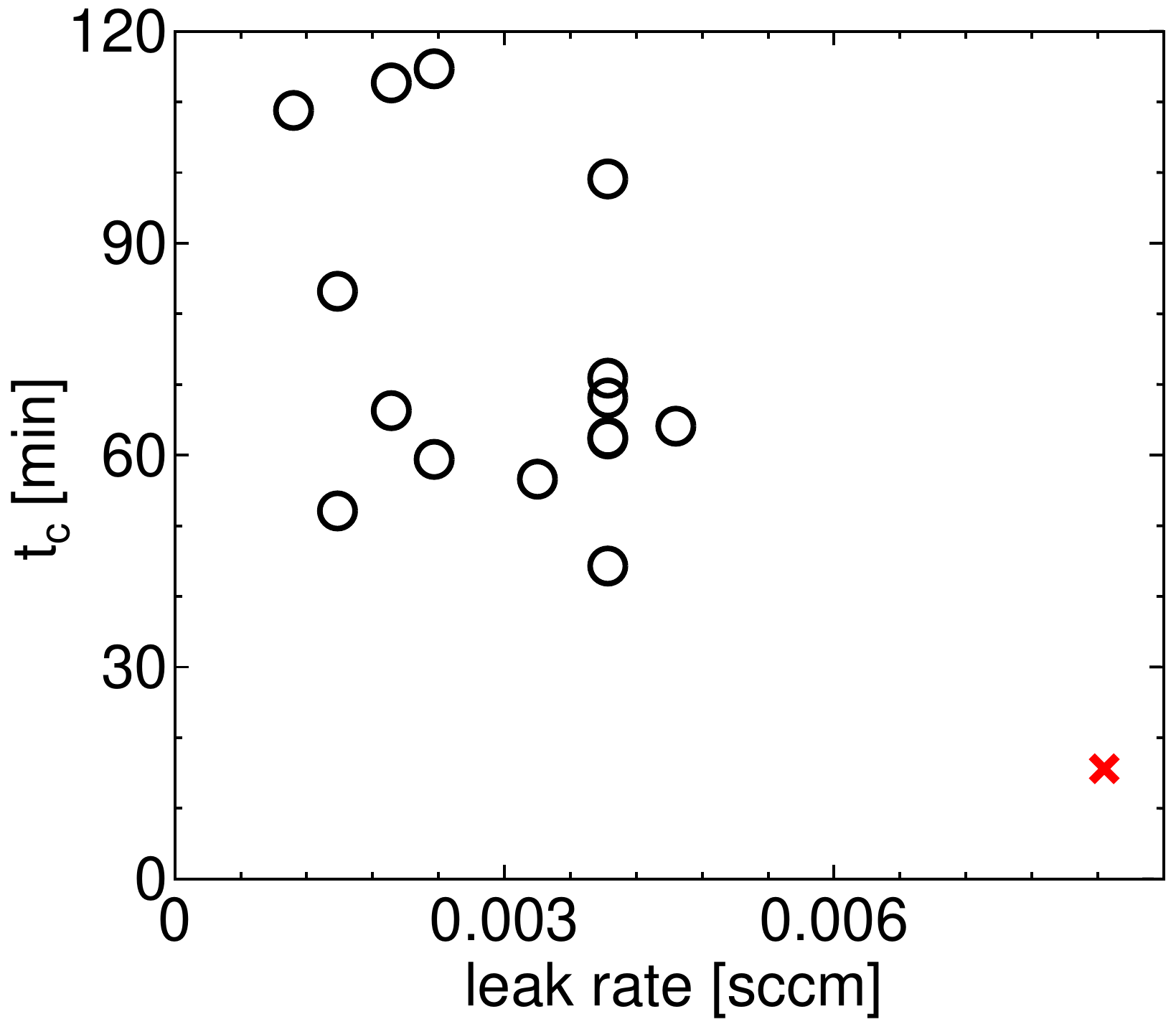}
		\caption{\label{t_c} Values of $t_{\mathrm{c}}$ for various leak rates determined from fitting absorbance data with $\beta\!>\!0$. In most cases, $t_{\mathrm{c}}\,\gg\,t_{\mathrm{CVD}}$, but the `$\times$' (lower-right) corresponds to the fast leak case shown in Fig.~\ref{fit-burning}b, and $t_{\mathrm{c}}$ is on the order of the CVD time. \newline}
	\end{center}
\end{figure}

From Eq.~(\ref{film-growth-rate-beta}) we can see that burning will eventually dominate, and the growth rate will become negative. This transition occurs at some time $t\,=\,t_{\mathrm{c}}$, when $\gamma(t_{\mathrm{c}})\,=\,0$ and $\beta=\gamma_{0}\exp\bigl(\frac{-t_{\mathrm{c}}}{\tau}\bigr)$. We can then obtain an expression for $t_{\mathrm{c}}$, which is
\begin{equation}
	t_{\mathrm{c}} = - \tau \log \left( \frac{\beta}{\gamma_{0}} \right).
	\label{crit-time}
\end{equation}
Various fittings using Eq.~(\ref{withbeta}) yielded values of $t_{\mathrm{c}}$ between 50 and 100 min, with an average of 70 min (Fig.~\ref{t_c}). 
Since this is considerably longer than the effective CVD time, burning is insignificant in most cases. However, when the leak rate is appreciable, $t_{\mathrm{c}}$ is comparable to the CVD time and burning of the film can become significant. For the two cases shown in Fig.~\ref{fit-burning}a, the values of $t_{\mathrm{c}}$ for the upper and lower cases are 102 and 15 min, respectively. Although accurate determination of $t_{\mathrm{c}}$ is difficult because in most cases $\beta~\ll~\gamma_{0}$, this is not important so long as $t_{\mathrm{c}}~\gg~t_{\mathrm{CVD}}$. Figure \ref{t_c} does not show any transition where burning effects emerge, but we presume it is safe to say that a leak rate below 0.005\,sccm should be sufficiently slow for burning to be negligible. It should be noted that a slow leak rate could be further slowed by flushing the system with Ar/H$_{\mathrm{2}}$ for one hour or more, but faster leak rates (e.g.~the case in Fig.~\ref{fit-burning}b) were unaffected by this treatment. This suggests the slow `leak' determined for the majority of cases may be due to outgassing from within the CVD chamber, rather than an actual leak in the vacuum system. Faster leak rates may be due to actual leaking of air into the chamber, causing the observed burning.

\subsection{Optical measurement of VA-SWCNT burning}
\begin{figure}[t]
	\begin{center}
		\includegraphics[width=0.48\textwidth]{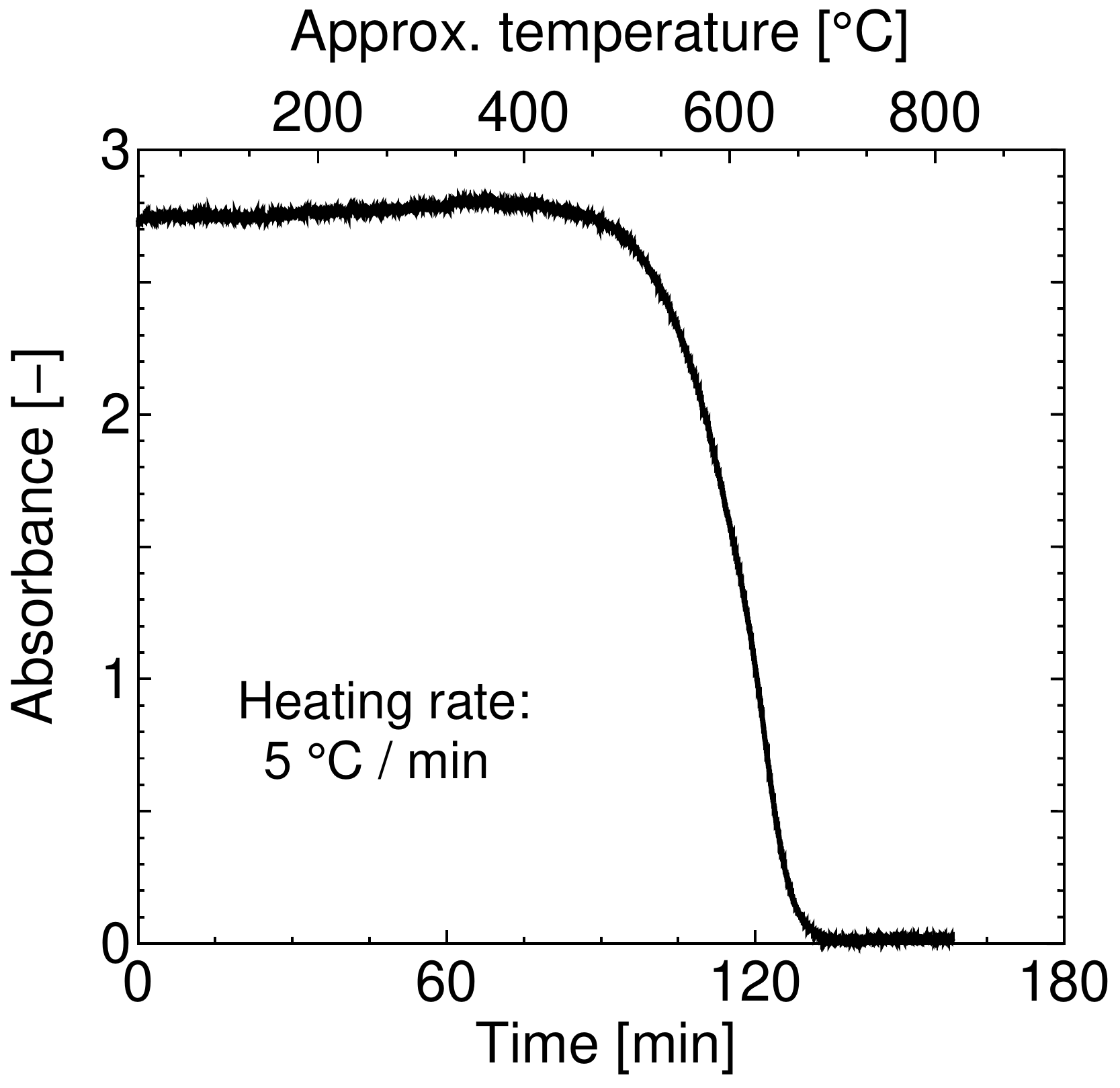}
		\caption{\label{TOA} Burning temperature of a VA-SWCNT film determined from {\em in situ\/} optical absorbance. Sample environment is dry air at 1\,atm, and heating rate was 5\,\textcelsius\//min. \newline}
	\end{center}
\end{figure}

The standard method of measuring the burning temperature of a material, thermogravimetric analysis (TGA), usually requires several hundred $\mu$g of material to obtain a decent measurement. This is usually a trivial amount, but can be significant for scarce or low-mass materials. Here we present a way of using the {\em in situ\/} optical measurement to determine the burning temperature of a VA-SWCNT film, which is difficult to measure by conventional TGA. In this measurement, shown in Fig.~\ref{TOA}, the optical absorbance of a VA-SWCNT film was measured while heating the film from room temperature to 925\,\textcelsius\/ at a constant rate of 5\,\textcelsius/min. The heating environment was dry air at 1\,atm. 
The upper axis shows the temperature inside the chamber, and the lower axis shows the heating time. Burning of the VA-SWCNTs is clearly seen by the significant decrease in absorbance at $\sim$\,600\,\textcelsius. This measurement is made over the cross-sectional area of the incident laser beam (a few $\mu$m$^{2}$), thus can be performed on a small, representative portion of a sample. Furthermore, using the optical method allows VA-SWCNTs to be measured in their as-grown state, whereas TGA requires some dispersion and collection of the SWCNTs, possibly from a number of different batches.

\section{Summary}
We investigated the growth dynamics of vertically aligned single-walled carbon nanotubes (VA-SWCNTs) using an {\em in situ\/} optical absorbance measurement. The growth can be described by an exponentially decaying growth rate and a reaction time constant. 
The temperature-pressure dependence of the VA-SWCNT film growth was investigated, and we find there is an optimum pressure at which the growth is maximized. 
Below this optimum pressure, which increases with CVD temperature, the growth rate is linearly dependent on ethanol pressure, indicative of a first-order reaction. In this pressure regime the reaction is rate-limited by the rate of arrival of ethanol at active catalyst sites. An Arrhenius plot of the initial growth rates corresponding to the optimum ethanol pressures yields an activation energy of approximately 1.5\,eV. 
A novel method for measuring the burning temperature of low-mass materials was also introduced, which combines the sensitivity of the {\em in situ\/} optical absorbance measurement with controlled oxidation of the material. This method may be attractive as an alternative to conventional methods when dealing with materials such as VA-SWCNTs, where the available quantity is limited, or for testing a small, representative portion of a larger sample.

\begin{ack}
The authors thank H.~M.~Duong at UT for experimental contributions. Part of this work was financially supported by KAKENHI \#19206024 and \#19054003 from JSPS, and SCOPE \#051403009 from MIC.
\end{ack}


\clearpage
\begin{thebibliography}{99}

\bibitem{Saito-text}
Saito R, Dresselhaus G, Dresselhaus MS.
\newblock Physical Properties of Carbon Nanotubes.
\newblock Imperial College Press, London; 1998.

\bibitem{Dresselhaus-text}
Dresselhaus MS, Dresselhaus G, Avouris Ph, editors.
\newblock Carbon Nanotubes: Synthesis, Structure, Properties, and Applications.
  vol.~80.
\newblock Springer, Berlin; 2001.

\bibitem{Baughman-Science297}
Baughman RH, Zakhidov AA, de~Heer WA.
\newblock Carbon nanotubes---{T}he route toward applications.
\newblock Science. 2002;297:787--792.

\bibitem{Murakami-CPL385}
Murakami Y, Chiashi S, Miyauchi Y, Hu M, Ogura M, Okubo T, et~al.
\newblock Growth of vertically aligned single-walled carbon nanotube films on
  quartz substrates and their optical anisotropy.
\newblock Chem Phys Lett. 2004;385:298--303.

\bibitem{Murakami-CPL377}
Murakami Y, Miyauchi Y, Chiashi S, Maruyama S.
\newblock Direct synthesis of high-quality single-walled carbon nanotubes on
  silicon and quartz substrates.
\newblock Chem Phys Lett. 2003;377:49--54.

\bibitem{Maruyama-CPL360}
Maruyama S, Kojima R, Miyauchi Y, Chiashi S, Kohno M.
\newblock Low-temperature synthesis of high-purity single-walled carbon
  nanotubes from alcohol.
\newblock Chem Phys Lett. 2002;360:229--234.

\bibitem{Murakami-CPL374}
Murakami Y, Miyauchi Y, Chiashi S, Maruyama S.
\newblock Characterization of single-walled carbon nanotubes catalytically
  synthesized from alcohol.
\newblock Chem Phys Lett. 2003;374:53--58.

\bibitem{Hata-supergrowth}
Hata K, Futaba DN, Mizuno K, Namai T, Yumura M, Iijima S.
\newblock Water-assisted highly efficient synthesis of impurity-free
  single-walled carbon nanotubes.
\newblock Science. 2004;306:1362--1364.

\bibitem{Noda-supergrowth-catalyst-support}
Noda S, Hasegawa K, Sugime H, Kakehi K, Zhang Z, Maruyama S, et~al.
\newblock Millimeter-thick single-walled carbon nanotube forests: {H}idden role
  of catalyst support.
\newblock Jpn J~Appl Phys. 2007;46:L399--L401.

\bibitem{Zhang-Dai-roles-H2-O2}
Zhang G, Mann D, Zhang L, Javey A, Li Y, Yenilmez E, et~al.
\newblock Ultra-high-yield growth of vertical single-walled carbon nanotubes:
  {H}idden roles of hydrogen and oxygen.
\newblock Proc Nat Acad Sci. 2005;102:16141--16145.

\bibitem{Zhong-VASWNTs-plasmaCVD}
Zhong GF, Iwasaki T, Honda K, Furukawa Y, Ohdomari I, Kawarada H.
\newblock Very high yield of vertically aligned single-walled carbon nanotubes
  by point-arc microwave plasma {CVD}.
\newblock Chem Vap Dep. 2005;11:127--130.

\bibitem{Eres-molecular-beam-VASWNTs}
Eres G, Kinkhabwala AA, Cui H, Geohegan DB, Puretzky AA, Lowndes DH.
\newblock Molecular beam-controlled nucleation and growth of vertically aligned
  single-wall carbon nanotube arrays.
\newblock J~Phys Chem~B. 2005;109:16684--16694.

\bibitem{Xu-Hauge-hot_filament_VASWNTs}
Xu YQ, Flor E, Kim MJ, Behrang H, Schmidt H, Smalley RE, et~al.
\newblock Vertical array growth of small diameter single-walled carbon
  nanotubes.
\newblock J~Am Chem Soc. 2006;128:6560--6561.

\bibitem{Noda-Co-Mo_combinatorial}
Noda S, Sugime H, Osawa T, Yoshiko T, Chiashi S, Murakami Y, et~al.
\newblock A simple combinatorial method to discover {C}o-{M}o binary catalysts
  that grow vertically aligned single-walled carbon nanotubes.
\newblock Carbon. 2006;44:1414--1419.

\bibitem{Zhang-Resasco-catal_density}
Zhang L, Tan Y, Resasco DE.
\newblock Controlling the growth of vertically oriented single-walled carbon
  nanotubes by varying the density of {C}o-{M}o catalyst particles.
\newblock Chem Phys Lett. 2006;422:198--203.

\bibitem{Maruyama-CPL403}
Maruyama S, Einarsson E, Murakami Y, Edamura T.
\newblock Growth process of vertically aligned single-walled carbon nanotubes.
\newblock Chem Phys Lett. 2005;403:320--323.

\bibitem{Murakami-PRL94}
Murakami Y, Einarsson E, Edamura T, Maruyama S.
\newblock Polarization dependence of the optical absorption of single-walled
  carbon nanotubes.
\newblock Phys Rev Lett. 2005;94:087402.

\bibitem{Murakami-Carbon43}
Murakami Y, Einarsson E, Edamura T, Maruyama S.
\newblock Polarization dependent optical absorption properties of single-walled
  carbon nanotubes and methodology for the evaluation of their morphology.
\newblock Carbon. 2005;43:2664--2676.

\bibitem{Dresselhaus-Eklund-Phonons}
Dresselhaus MS, Eklund PC.
\newblock Phonons in carbon nanotubes.
\newblock Adv Phys. 2000;49:705--814.

\bibitem{Murakami-PRB71}
Murakami Y, Chiashi S, Einarsson E, Maruyama S.
\newblock Polarization dependence of resonant {R}aman scatterings from
  vertically aligned {SWNT} films.
\newblock Phys Rev~B. 2005;71:085403.

\bibitem{XiangRong-13C_root_growth}
Xiang R, Zhang Z, Ogura J, Okawa J, Einarsson E, Miyauchi Y, Shiomi J, Maruyama S.
\newblock Vertically Aligned $~{13}${C} single-walled carbon nanotubes from no-flow alcohol chemical vapor deposition and their root growth mechanism.
\newblock Jpn J~Appl Phys. 2008; in press.

\bibitem{Hu-J_Catal}
Hu M, Murakami Y, Ogura M, Maruyama S, Okubo T.
\newblock Morphology and chemical state of {C}o-{M}o catalysts for growth of
  single-walled carbon nanotubes vertically aligned on quartz substrates.
\newblock J~Catal. 2004;225:230--239.

\bibitem{XiangRong-diffusion_limit}
Xiang R, Yang Z, Zhang Q, Luo G, Qian W, Wei F, Einarsson E, Maruyama S.
\newblock Growth deceleration of vertically aligned carbon nanotube arrays: {C}atalyst deactivation or feedstock diffusion controlled?
\newblock Submitted to J Phys Chem C.

\bibitem{Puretzky-kinetic_model}
Puretzky AA, Geohegan DB, Jesse S, Ivanov IN, Eres G.
\newblock In situ measurements and modeling of carbon nanotube array growth kinetics during chemical vapor deposition.
\newblock Appl Phys~A 2005;81:223-240.

\bibitem{Futaba-supergrowth_kinetics}
Futaba DN, Hata K, Yamada T, Mizuno K, Yumura M, Iijima S.
\newblock Kinetics of water-assisted single-walled carbon nanotube synthesis
  revealed by a time-evolution analysis.
\newblock Phys Rev Lett. 2005;95:056104.

\bibitem{Louchev-CVD-diffusion-kinetics}
Louchev OA, Laude T, Sato Y, Kanda H.
\newblock Diffusion-controlled kinetics of carbon nanotube forest growth by chemical vapor deposition.
\newblock J Chem Phys. 2003;118:7622-7634.

\bibitem{Hai-Thermal_degradation}
Duong HM, Einarsson E, Okawa J, Xiang R, Maruyama S.
\newblock Thermal degradation of single-walled carbon nanotubes.
\newblock Jpn J~Appl Phys. 2008; in press.

\end{thebibliography}


\end{document}